# Design of efficient high-order immersed metagratings using an evolutionary algorithm

Dhwanil Patel,[1, 2] Jacob de Nobel,[3] Anna V. Kononova,[3] Bernhard R. Brandl,[1, 4] Ralf Kohlhaas,[2, 5, *]

[1] *Leiden Observatory, Leiden University, Leiden, The Netherlands*
[2] *SRON Netherlands Institute for Space Research, Leiden, The Netherlands*
[3] *LIACS, Leiden University, The Netherlands*
[4] *Faculty of Aerospace Engineering, Delft University of Technology, Kluyverweg 1, 2629 HS Delft, The Netherlands*
[5] *Department of Precision and Microsystems Engineering, Delft University of Technology, 2628 CD Delft, The Netherlands*
[*] *r.kohlhaas@tudelft.nl*

**Abstract:** Immersed reflection gratings improve spectral resolving power by enabling diffraction within a high refractive index medium. This principle has been widely adopted to make grating spectrometers more compact. Conventional immersed gratings have blazed profiles which typically show the highest efficiency for one main design wavelength. In addition, the blazed profiles tend to cause significant polarization sensitivity. In this work, we propose an alternative approach for designing an immersed grating composed of sub-wavelength structures, designed to increase diffraction efficiency and reduce polarization dependence. For a theoretical demonstration, a reflective metagrating immersed in silicon is optimized over the short-wave infrared band-3 (SWIR-3, here $2.304\,\mu$m–$2.405\,\mu$m), targeting the same diffraction angles as the immersion grating used in the Sentinel-5 Earth observation mission. The structure is optimized using a modified Covariance Matrix Adaptation Evolution Strategy (CMA-ES). The optimized immersed metagrating achieves an average efficiency of (over the SWIR-3 band) $\sim 78\%$, compared to $\sim 62\%$ for the conventional immersed blazed grating, and reduces polarization sensitivity from roughly $\sim 15\%$ to $\sim 5\%$. A manufacturing tolerance analysis is also conducted to evaluate the design's performance under systematic manufacturing errors, which revealed a degradation of $\sim 10\%$ efficiency at feature size errors of $\pm 25$ nm and almost negligible effect on the efficiency at $-10$ nm and of $\sim 5\%$ at $+10$ nm.

## 1. Introduction

High-resolution spectrometers are essential in ground-based and space-based astronomy, where compact and high-performance optical components are critical. Immersed diffraction gratings are particularly favourable due to their ability to provide high resolving power while maintaining a compact volume. They are currently used in instruments such as SPIFFIER on the VLT [1], and are planned for first-light instruments on upcoming extremely large telescopes, such as METIS for the Extremely Large Telescope (ELT) [2] and GMTNIRS for the Giant Magellan Telescope (GMT) [3], as well as in space-based missions like Sentinel-5 [4].

Blazed gratings are one of the most commonly used diffraction grating types in these applications, due to their ability to efficiently direct light into a selected diffraction order. A blazed grating concentrates power into a specific diffraction order $m$ corresponding to a diffraction angle $\theta_d$. This behaviour is governed by the grating equation $\sin(\theta_i) + \sin(\theta_d) = m\,(\lambda_{\text{eff}}/P)$, where $\theta_i$ is the incidence angle, $P$ is the period, and $\lambda_{\text{eff}}$ is the effective wavelength. This is the blazing wavelength for which the blazed grating is configured to work. The resolving power of such a grating is given by $\mathcal{R} = mN$, where $N$ is the number of grating lines. Based on this, the diffraction order needs to be increased to enhance the resolving power while maintaining a compact grating (low $N$). Hence, to design a higher-order grating, an immersion medium with refractive index $n$ can be used to modulate $\lambda_{\text{eff}} = \lambda/n$. The resolving power of the grating would then scale with $n$ of the immersion material.

The power concentrated into the diffraction angle ($\theta_d$) depends on the groove shape and the incidence angle ($\theta_i$) [5]. Specifically, the blaze angle $\theta_b$ of the grating (see Figure 1 (a) and (b)) is related to the diffraction and incidence angles by the blazing condition. This is given by $\theta_b = \theta_i \pm \theta_d$. A typical strategy to obtain a blazed grating profile is through wet etching into a crystalline material, where the resulting etching angle (formed by the two groove surfaces, $\eta$ in Figure 1 (b)) is fixed by the crystal axes of the material [6]. For example, in KOH (potassium hydroxide) etching of amorphous silicon (a-Si), the etching angle $\eta$ is fixed to $\sim 70.5°$ [4]. This limitation affects the efficiency of the grating since an idealized groove shape is not realized. This can be mitigated using a more precise machining process, such as single-point diamond turning (SPDT) machining as used for METIS's mid-infrared Ge grating [2]. However, SPDT struggles with shorter periods (< 5-10 $\mu$m), which is required for optical, near-infrared and infrared applications.

Metasurfaces are used for robust wavefront control [7] and have been applied in areas such as beam steering [8], waveplates [9], and diffraction gratings operating in both transmission [10] and reflection [11]. Blazed diffraction metagratings based on binary surfaces, designed to mimic the blazed $2\pi$ phase profile, have also been demonstrated in optical [12, 13], infrared [14], and even broadband [15] applications. Beyond conventional diffraction control, metagratings have also been used to generate structured beams such as vortex beams [16], and for polarisation state generation [17].

Metasurfaces are implemented using lithography techniques in the form of nanopillars [14], ridges [13] [18], or both [19]. By varying dimensions (see Figure 1 (d)) the subwavelength structure's ability to behave as an effective homogeneous medium [20] can be leveraged. This is referred to as effective index dispersion engineering based on Rytov's effective medium theory (EMT) [21]. A blazed profile, even when ideally implemented using metasurfaces, offers only a single degree of freedom (the blaze angle that is mimicked by the metasurfaces). This limits the ability to design higher-order gratings with minimal polarization dependence, even over narrowband spectral ranges. Additionally, as evident from the grating equation, dispersion in blazed gratings is inherently chromatic, and the blaze angle is optimized to concentrate power into a specific diffraction order at an operating wavelength. This leads to wavelength and polarization-dependent variations in diffraction efficiency. Therefore, rather than using metasurfaces strictly to emulate an idealized sawtooth blazed profile ("Blazed profile" and "sawtooth profile" are used interchangeably in this paper) for one wavelength, they may be employed more flexibly to optimize the grating's performance directly over a wavelength range. Additionally, the modulation by EMT depends on the incident field orientation and is therefore polarization sensitive [22]. Hence, as with conventional blazed gratings [23], metagratings also exhibit polarization dependence. By directly optimizing the performance of the metagrating, the sensitivity to polarization and chromaticity can be minimized compared to a conventional sawtooth grating, as shown in [24] for a $\pm$1st order grating case.

In this study, an evolutionary algorithm (EA) described in [25] is used to optimize the design of a higher-order immersed metagrating operating in the shortwave infrared (SWIR-3) band ($2.304\,\mu$m – $2.405\,\mu$m). The reference grating for the Sentinel-5 mission was manufactured with KOH wet etching, which leads to smooth side walls, but constrains the groove angle [26]. Similar to the Sentinel-5 grating, the metagrating is immersed in silicon ($Si$) and coated with aluminum ($Al$) to operate in reflection, but uses nanopillars for the grating operation. Figure 1 compares schematics of such a grating with a conventional blazed sawtooth grating. Unlike the conventional approach that mimics a single blaze angle, the metagrating is directly optimized over a wavelength range to maximize diffraction efficiency and minimize polarization sensitivity. The motivation behind using an optimization algorithm and the principles behind the design are explained in Section 2.1. The design problem, the optimization algorithm, the figure of merit and the simulation method are described in Section 2.2. Results from the optimization process and the optimal metagrating design are presented in subsection 3.1. The performance of the optimized design—including its diffraction efficiency, polarization dependence, and dispersion characteristics—is shown in Section 3.2. The limitations of the current optimization approach and the manufacturing considerations are also discussed in Sections 3.1 and 3.2, respectively. Finally, the limitations of this study and directions for future work with a summary are outlined in Section 4.

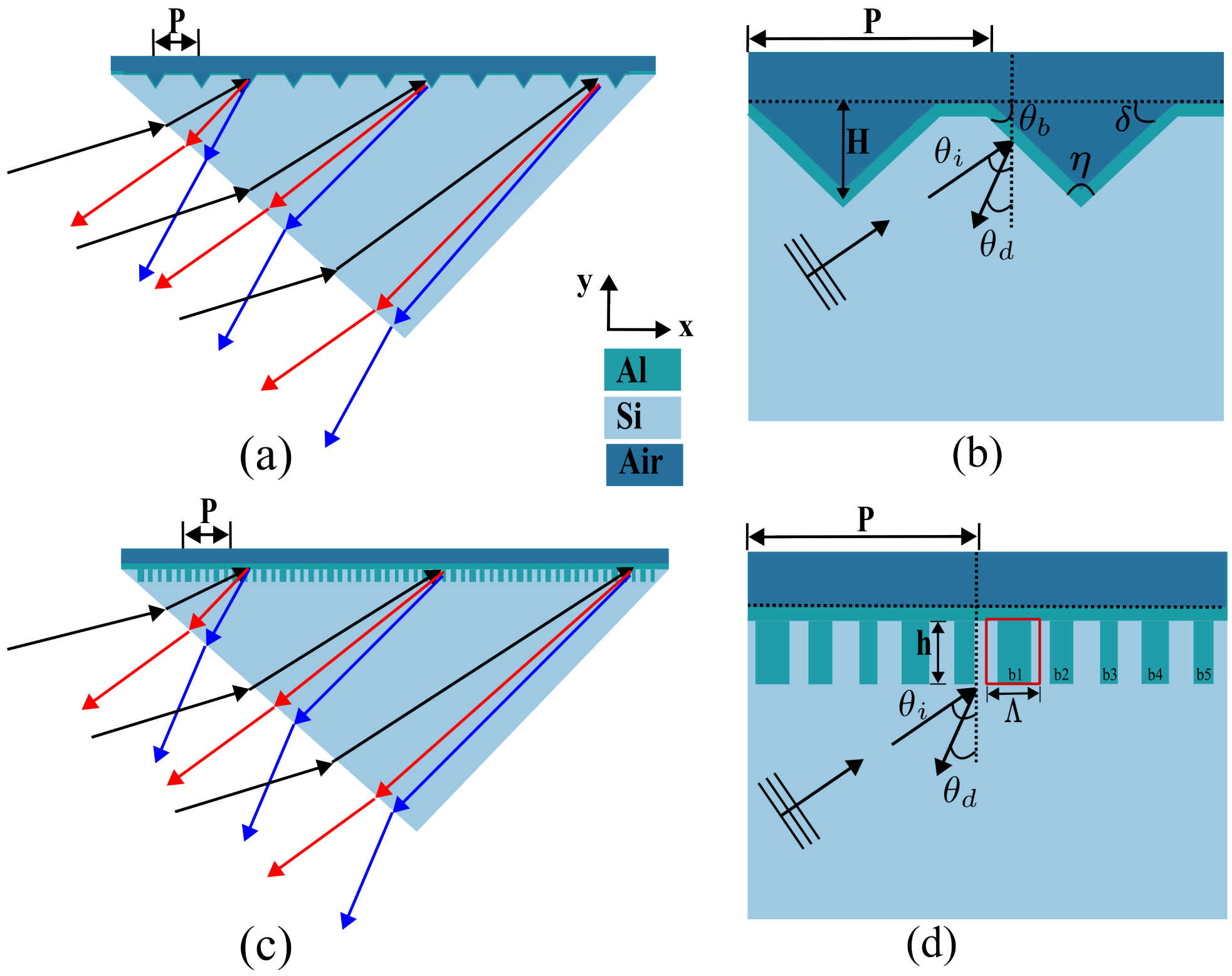

Fig. 1. Illustration of an immersed grating with sawtooth (or triangular) blaze profile in (a) and an immersed metagrating based on subwavelength pillars in (c). A polychromatic incident ray (black) is diffracted into the blue and red rays representing the shorter and longer wavelengths of the light, respectively.A detailed schematic of the profiles with two unit cells each of a sawtooth and metagrating is shown in (b) and (d), respectively. The unit cells have Period $P$ and the light with incidence angle $\theta_i$ gets diffracted at angle $\theta_d$. The sawtooth profile is defined with 3 degrees of freedom, namely height, etching angle $\eta$ and blaze angle $\theta_b$. The metagrating profile has multiple subwavelength structures in one period P with a lattice constant $\Lambda$. This is highlighted by the red box in plot (d). Each lattice then had structures with varying fill factor ($b_i$). This fill factor can vary in both the XY plane and XZ plane. The metasurfaces have the height $h$ in the y dimension.

## 2. Theory and methods

### 2.1. Design principle

The design is constrained to have a fixed period $P$ (see Figure 1) of 2.07 $\mu$m, with an incidence angle $\theta_i = 62.6°$, diffraction order $m = -5$, and consequently a diffraction angle $\theta_d = -49.8°$. These constraints are based on the design parameters of the Sentinel-5 SWIR-3 (centered at $\lambda_{\text{mean}} = 2.345\,\mu$m with $\sim 4\%$ bandwidth) blazed grating described in [4]. This establishes a baseline for comparing the proposed immersed metagrating to a conventional immersed blazed grating.

The metagrating, in its simplest form, consists of a periodically arranged subwavelength structure. For a structure with period $\Lambda$ composed of alternating materials with complex refractive indices $\tilde{n}_1$ and $\tilde{n}_2$, and respective widths $a$ and $b$, operating at a free-space wavelength $\lambda$, the effective refractive index $n_{\text{eff}}$ is governed by the full formalism of Rytov's theory. The expressions for the two orthogonal polarizations are given by Equations 1 and 2, respectively [21,22].

$$\begin{aligned}&\sqrt{\tilde{n}_2^2 - \left(n_{\text{TE}}^{\text{EMT}}\right)^2}\tan\left[\frac{\pi\Lambda}{\lambda}\left(\frac{b}{\Lambda}\right)\sqrt{\tilde{n}_2^2 - \left(n_{\text{TE}}^{\text{EMT}}\right)^2}\right]\\ &= -\sqrt{\tilde{n}_1^2 - \left(n_{\text{TE}}^{\text{EMT}}\right)^2}\tan\left[\frac{\pi\Lambda}{\lambda}\left(\frac{a}{\Lambda}\right)\sqrt{\tilde{n}_1^2 - \left(n_{\text{TE}}^{\text{EMT}}\right)^2}\right]\end{aligned} \tag{1}$$

$$\begin{aligned}&\frac{\sqrt{n_2^2 - \left(n_{\text{TM}}^{\text{EMT}}\right)^2}}{n_2^2}\tan\left[\frac{\pi\Lambda}{\lambda}\left(\frac{b}{\Lambda}\right)\sqrt{n_2^2 - \left(n_{\text{TM}}^{\text{EMT}}\right)^2}\right]\\ &= -\frac{\sqrt{n_1^2 - \left(n_{\text{TM}}^{\text{EMT}}\right)^2}}{n_1^2}\tan\left[\frac{\pi\Lambda}{\lambda}\left(\frac{a}{\Lambda}\right)\sqrt{n_1^2 - \left(n_{\text{TM}}^{\text{EMT}}\right)^2}\right]\end{aligned} \tag{2}$$

Because the transcendental EMT equations (Equations 1–2) involve tangent functions with periodic singularities, they admit an infinite family of solutions for $n_{\text{TE}}^{\text{EMT}}$ and $n_{\text{TM}}^{\text{EMT}}$ that depend on the materials ($n_1$, $n_2$), wavelength($\lambda$) and the design parameters($\Lambda$, $a$, $b$). In practice, only the principal (first-order) root can be retained [20,22], but is only valid when the grating period $\Lambda$ is vanishingly small compared to the free-space wavelength $\lambda$. In our design, the period 2.07 $\mu$m is divided into five sub-periods of 0.414 $\mu$m to meet fabrication aspect-ratio constraints; although 0.414 $\mu$m $< 2.345\,\mu$m, the ratio ($\Lambda/\lambda \approx 0.18$) is large enough for the first order-approximation to hold. Additionally, the nonzero extinction coefficients of silicon and aluminum would shift the roots. Consequently, there is no closed-form guarantee that the simple first-order EMT solution will maximize the grating efficiency (defined here as diffraction efficiency $\times$ reflection efficiency). Moreover, to use the Rytov-based EMT to approximate a profile, an ideal profile is required that performs well across the spectral range and accounts for polarization effects. A conventional blazed (sawtooth) profile is typically optimized at a single wavelength within the spectral band, and its performance is therefore not uniform across the band. Furthermore, the blazed profile does not account for polarization. To address these limitations and to exploit subwavelength structuring to engineer an arbitrary effective medium, an evolutionary algorithm (EA) is employed. The EA effectively optimizes the geometry to maximize the overall grating efficiency while also minimizing sensitivity to polarization and wavelength variations through an appropriately defined figure of merit (FOM). The underlying principle enabling this approach is that the optimizer discovers metasurfaces with a refractive index profile that is well adapted to the problem at hand.

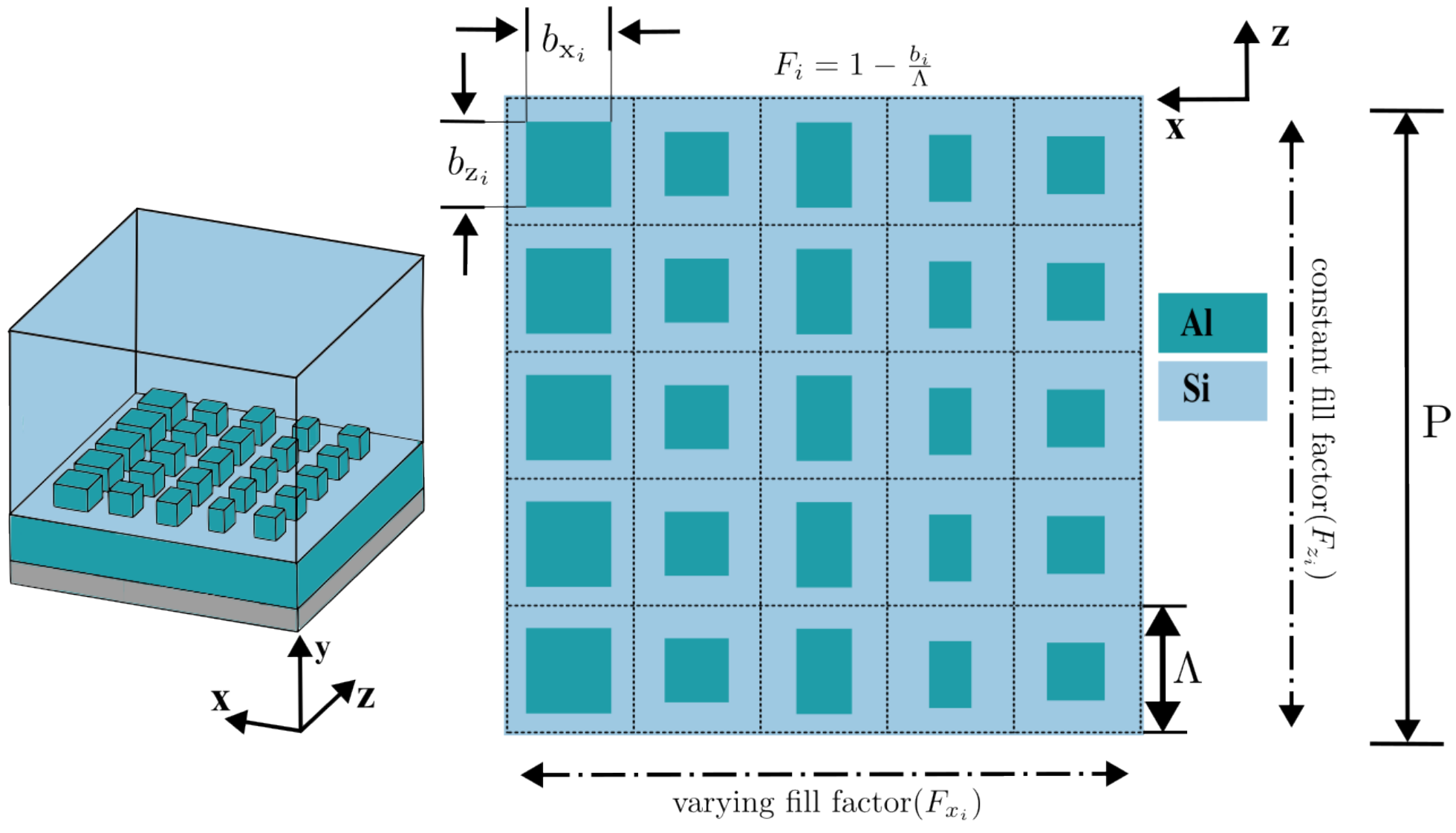

Fig. 2. A cartoon of the immersed metagrating unit cell of period PxP is shown on the left with the xyz planes labelled. On the right, the XZ plane of the design problem is shown. It is 5x5 lattice grid with lattice constant $\Lambda = 0.414\mu$m within a grating period of $P = 2.070\mu$m. The center of each lattice and its pillar coincide. The pillars have the same height and varying fill factors ($F_{x_i}$ and $F_{z_i}$) in the x and z dimensions. The fill factor of the pillars changes in the x dimension, but is constant in the z dimension.

### *2.2. Simulation and optimization of design*

To employ an EA, the constants, the figure of merit and the design variables are first defined. A schematic of the design space is shown in Figure 2. The overall grating period is $P = 2.07\ \mu$m, which is subdivided into five equal segments to obtain a subwavelength period $\Lambda = P/5 = 0.414\ \mu$m. This choice ensures the structures are within the subwavelength regime ($\Lambda < \lambda$) and also maintains each feature well above 100 nm for compatibility with manufacturing. Furthermore, an Al layer of 400 nm is placed on top (along the $y$-direction), as shown in Figure 1(d). This thickness was chosen as the minimum value that still preserves the true reflection efficiency of Al, while avoiding any increase in simulation time.

To minimize polarization effects and ensure optimized performance across the full SWIR-3 band, the figure of merit (FOM) is defined as the average grating efficiency $\eta$ at diagonal (D) polarization across three representative wavelengths: the minimum ($\lambda_1$), mean ($\lambda_2$), and maximum ($\lambda_3$) within the band. The D polarization corresponds to a polarization angle of 45°, meaning the electric field is oriented at 45° with respect to the plane of incidence. This makes it an equal linear combination of the two orthogonal polarization states: S polarization (90°) and P polarization (0°). Consequently, the grating efficiency under D polarization is the arithmetic mean of the efficiencies under S and P polarizations, effectively optimizing the structure's response to both the linear polarization states.

To evaluate the design, the finite-difference time-domain (FDTD) method is used, specifically the Lumerical FDTD:3D electromagnetic solver package (v.8.31.3683, Lumerical Inc.). Each unit cell with period ($P$) is simulated using Bloch-periodic conditions that compensate for the additional phase due to the oblique incidence angle ($\theta_i = 62.6°$). The grating efficiency is calculated by multiplying the diffraction efficiency of the structure at the angle corresponding to the -5th order (m) by the reflection efficiency of the grating. The simulations are performed at 14 points-per-wavelength (ppw) (corresponding to mesh level 3 in Lumerical FDTD) during optimization, and to gauge the performance of the optimized metagrating. The grating efficiencies ($\eta_{\mathrm{s}}$ and $\eta_{\mathrm{p}}$) of the optimized grating at the two linear polarizations (S and P (not to be confused with Period $P$) across the SWIR-3 band are calculated. Following that, the metagrating's polarization sensitivity ($\eta_{\mathrm{pol}}$) defined by equation 3 is determined.

$$\eta_{\mathrm{pol}} = \frac{\eta_{\mathrm{s}} - \eta_{\mathrm{p}}}{\eta_{\mathrm{s}} + \eta_{\mathrm{p}}} \tag{3}$$

The diffraction angle for the wavelength across the band is also calculated. The performance metrics of the metagrating are compared to the conventional blazed grating.

As for the design variables, each $P \times P$ unit cell of the grating is represented by a $5 \times 5$ grid of subcells. As mentioned previously, the $5 \times 5$ grid of subcells comes from the period divided into 5 parts, so as to balance the subwavelength regime and manufacturing tolerance. In the $x$–direction, these subcells vary in width and length (varying fill-factor) to approximate the target phase profile; in the $z$–direction, the pattern is kept uniform (constant fill-factor), yielding an array of nanopillars (refer to figure 2). This effectively ensures that the diffraction occurs only along one direction and only the fundamental order is excited in the other direction. Furthermore, the metasurfaces are limited to nanopillars rather than ridges. It has been shown before that using nanopillars rather than continuous ridges could improve the stability of the structure while keeping a low fill-factor [27]. All pillars share a common height $h$. The common height means the manufacturing is simplified, and the phase difference between each metasurface is purely from the fill-factor difference. Thus, the structure is parameterized by eleven degrees of freedom: the width ($b_{x_i}$) and length ($b_{z_i}$) of the 5 subcells (i = 1 - 5) along the $x$ direction, and the pillars' height ($h$) (again refer to figure 2). This yields the design vector $\mathbf{x} = [b_{x_1}, b_{z_1}, \ldots, b_{x_5}, b_{z_5}, H]$ of length $n_{\mathbf{x}} = 11$. A numeric optimizer $\mathbf{x}^*$ is found which maximizes our figure of merit (FOM) $f : \mathbf{S} \rightarrow \mathbb{R}$:

$$\mathbf{x}^* = \arg\max_{x \in \mathbf{S}} f(\mathbf{x}) \tag{4}$$

, where $\mathbf{S} \subseteq \mathbb{R}^n$ denotes our design space. Let $\eta_m(\mathbf{x}, \lambda)$ be the simulated efficiency for design $\mathbf{x}$ at wavelength $\lambda$ for diffraction order $m = 5$, then $f(\mathbf{x})$ is defined as:

$$f(\mathbf{x}) = \frac{1}{3} \sum_{i=1}^{3} \eta_m(\mathbf{x}, \lambda_i) \tag{5}$$

Since obtaining a closed-form analytical solution is infeasible, it is considered a black-box optimization problem [28]. As such, a heuristic optimization algorithm can be applied, which utilizes a stochastic process to optimize $\mathbf{x}$ by considering only the value of $f(\mathbf{x})$. Here, the Covariance Matrix Adaptation Evolution Strategy (CMA-ES) [29] is used. The CMA-ES is an EA designed for solving challenging continuous black-box optimization problems and is considered state-of-the-art in heuristic optimization [30]. The algorithm can be considered a second-order method, which estimates the parameters of a multivariate Gaussian distribution $\mathcal{N}(\mathbf{m}, \sigma^2\mathbf{C})$ that it uses to sample new $\mathbf{x}$. At every iteration, the algorithm samples $k$ candidate solutions of which $l < k$ are selected (Note that in standard CMA-ES literature $\mu$ and $\lambda$ are used for $l$ and $k$ resp.). The method then updates the parameters of the distribution, such that the likelihood of sampling the $l$ selected steps again is maximized. In essence, this creates a sampling distribution that forms a hyperellipsoid pointed towards the direction of most progress.

Over the years, several extensions to the CMA-ES have been proposed to enhance its search performance, many of which have been implemented within the Modular CMA-ES framework [31], which is utilized in this work. Specifically, the variant used here includes active covariance matrix updates [32] and mirrored sampling [33]. Since CMA-ES is defined as a *minimization* algorithm, the function $-f(\mathbf{x})$ is optimized internally. In this study, CMA-ES is applied as a practical optimization tool rather than investigated as an algorithmic contribution. It was chosen for its robustness in non-convex black-box problems, comparable to the present design task [34–36]. Other stochastic methods (e.g., PSO, DE, Bayesian optimization) could also be used, but a full benchmarking is beyond the scope of this work. CMA-ES serves here as a representative, well-established optimizer with demonstrated reliability [30]. The implementation follows the Modular CMA-ES framework (See: [31]) , with population size $k = 11$, parent number $l = 5$, initial step size $\sigma_0 = 0.6$, and a maximum of $N_{\text{eval}} = 500$ evaluations. The initial mean $\mathbf{m}_0$ was placed at the center of the design space with isotropic covariance $\mathbf{C}_0 = \mathbf{I}$. The algorithm employs the standard (non-elitist) selection scheme with weighted recombination, adhering to the canonical CMA-ES [29]. The total computational cost is dominated by the FDTD simulations, each of which takes ± 12 minutes per evaluation, while the optimizer's overhead is negligible. Because CMA-ES is a stochastic, derivative-free method, conventional convergence-rate metrics are not applicable; convergence is instead indicated by the stabilization of the objective value.

## 3. Results and discussion

### 3.1. CMA-ES optimization results

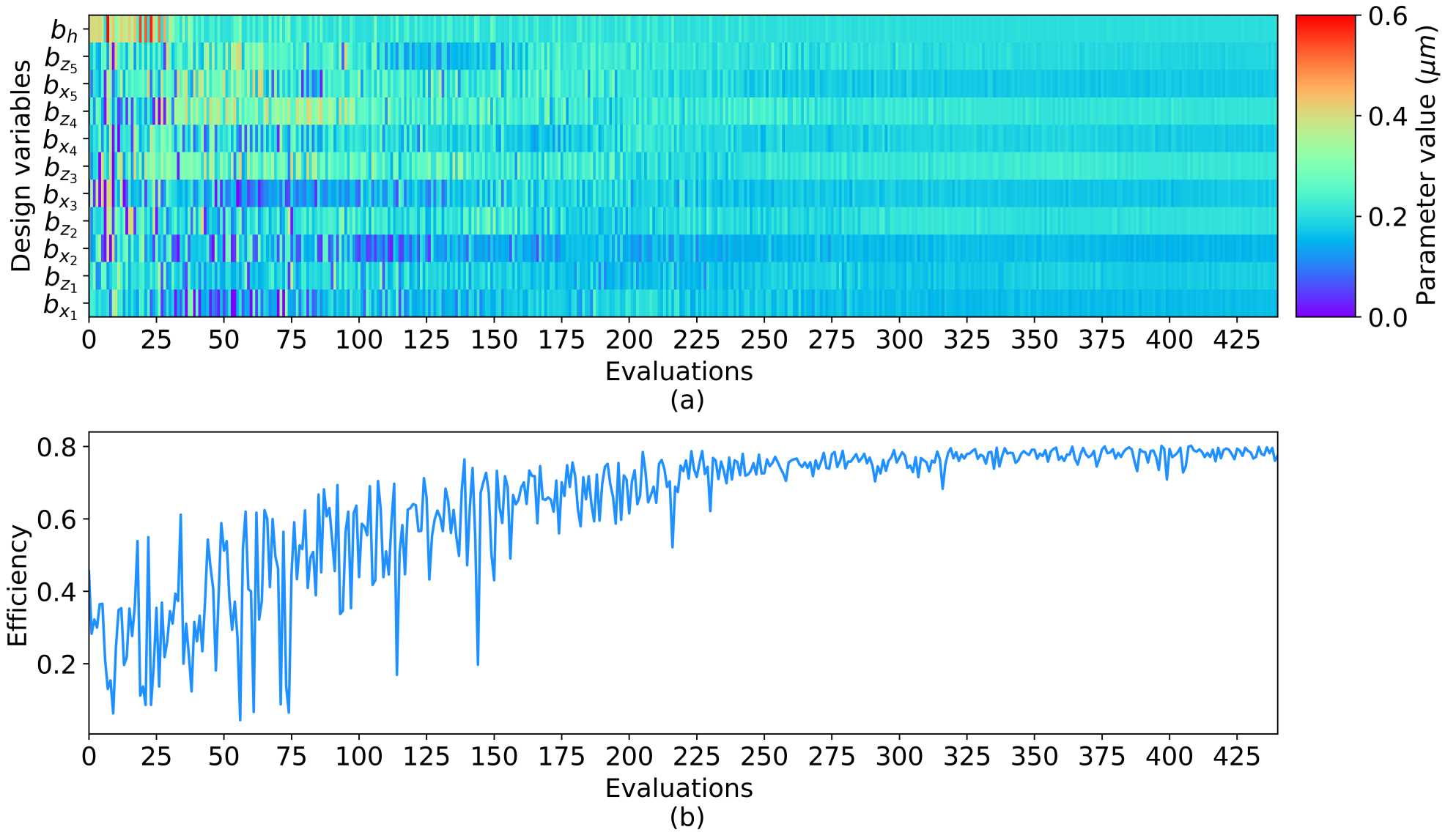

Fig. 3. Results from the CMA-ES optimization of the width $(x_i)$ and length $(z_i)$ of each of the 5 pillars and the height $H$, hence 11 degrees of freedom (DOF). The evolution of the DOF over each iteration of the optimization is given in (a), where the colorbar represents the value of each DOF in microns. The evolution of the FOM (averaged 5th order efficiency across the wavelengths) over the iteration is shown in (b).

The convergence behaviour of the CMA-ES optimization is illustrated in Figure 3. As shown, the algorithm begins with a relatively low-performing initial sample, characterized by a broad spread in the design parameter space. Over successive iterations, the population rapidly converges toward a region of significantly higher performance, corresponding to a local optimum in the objective landscape. Early in the optimization, the variance of the sampled solutions is large, reflecting the algorithm's exploratory phase as it searches the global design space. This variance decreases steadily as the algorithm homes in on a promising region, ultimately stabilizing around a consistent set of design parameters. This behaviour is typical of CMA-ES, as it adaptively adjusts its sampling distribution to match the local search landscape.

| **Parameter** | $b_{X_1}$ | $b_{Z_1}$ | $b_{X_2}$ | $b_{Z_2}$ | $b_{X_3}$ | $b_{Z_3}$ | $b_{X_4}$ | $b_{Z_4}$ | $b_{X_5}$ | $b_{Z_5}$ | H |
|---|---|---|---|---|---|---|---|---|---|---|---|
| **Value [$\mu$m]** | 0.163 | 0.209 | 0.179 | 0.171 | 0.152 | 0.211 | 0.208 | 0.174 | 0.177 | 0.185 | 0.200 |

Table 1. Optimized design parameters; widths $X_i$, $Z_i$ in x and z directions, and the height $H$ of the subwavelength structures.

Although minor improvements in the objective function may still be possible with additional iterations, the computational cost of the simulations is substantial. Therefore, further optimization was not pursued beyond the point of apparent convergence. Additionally, the precision of the design variables is limited by the capabilities of the fabrication process. Beyond a certain point, optimizing to sub-nanometer precision would yield no practical benefit, as such resolution cannot be reliably realized in manufacturing. The final optimized design parameters resulting from the CMA-ES procedure are given in Table 1, and are visualized in the unit cell in Figure 4.

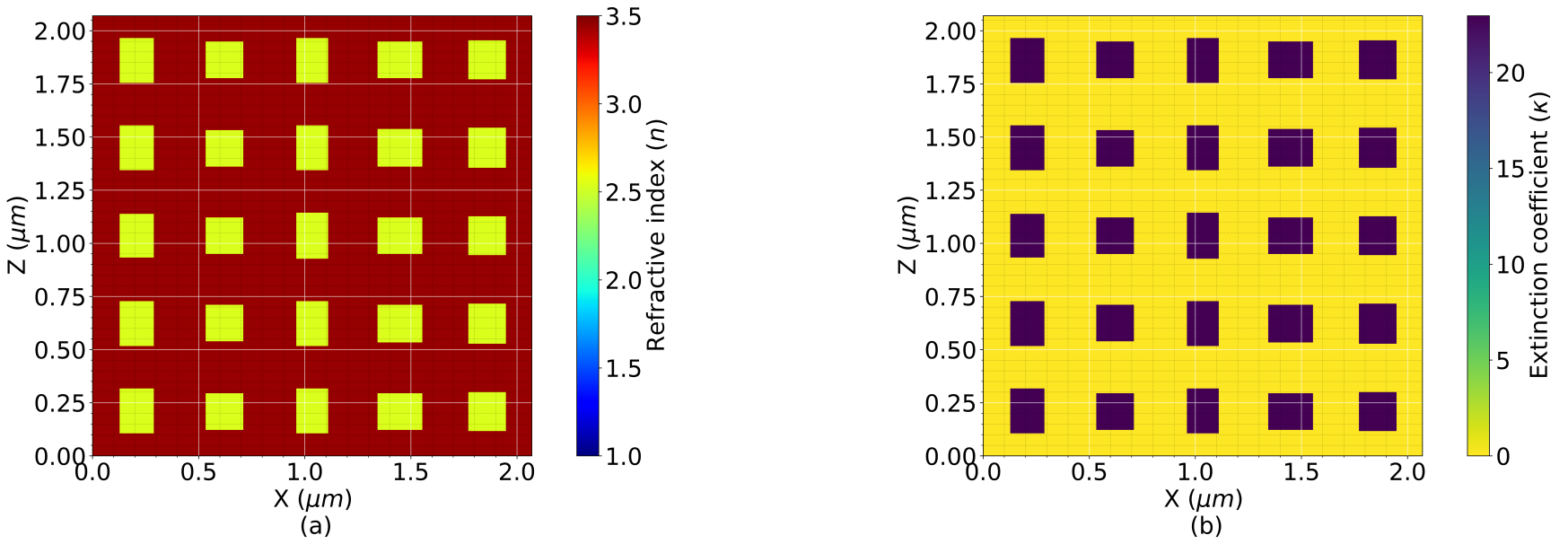

Fig. 4. Complex refractive index of the metagrating for XZ plane in (a) and (b). The two color bars represent the real refractive index ($n$) and the imaginary extinction coefficient ($\kappa$). The graph visualizes the optimal width and length of the metasurfaces.

All the pillars have feature sizes just above 150 nm, which is well within the current e-beam lithography capabilities for Si, as discussed in [37]. The designed metagrating can even be prototyped with deep UV photo-lithography (DUV-PL) with methods to improve for smaller feature sizes ($\sim$ 100 nm) at reduced precision as demonstrated in [38]. The largest aspect ratio is $\sim$ 1.3, which further relaxes the manufacturing constraints. The Al filling in the etched pillars and coating could be done by sputtering.

### *3.2. Performance of the optimized metagrating*

Figure 5 shows the grating efficiency over the SWIR-3 band at the −5th order. The efficiency over the complete wavelength range for both linear polarizations is $\gtrsim$ 74%, with efficiency for S polarization better than that of P. This is an improvement compared to conventional blazed immersed grating, where the maximum efficiency was just above $\sim$ 72%. Table 2 shows the average efficiency over the SWIR-3 band for both meta- and conventional blazed grating for all polarization states. The average efficiencies improved by $\sim$ 15% for all three polarizations. Hence, reducing chromaticity compared to a conventional blazed grating is minimized. This is also clearly evident from the Figure, where the slope for the metagrating curves compared to that of the sawtooth is much smaller.

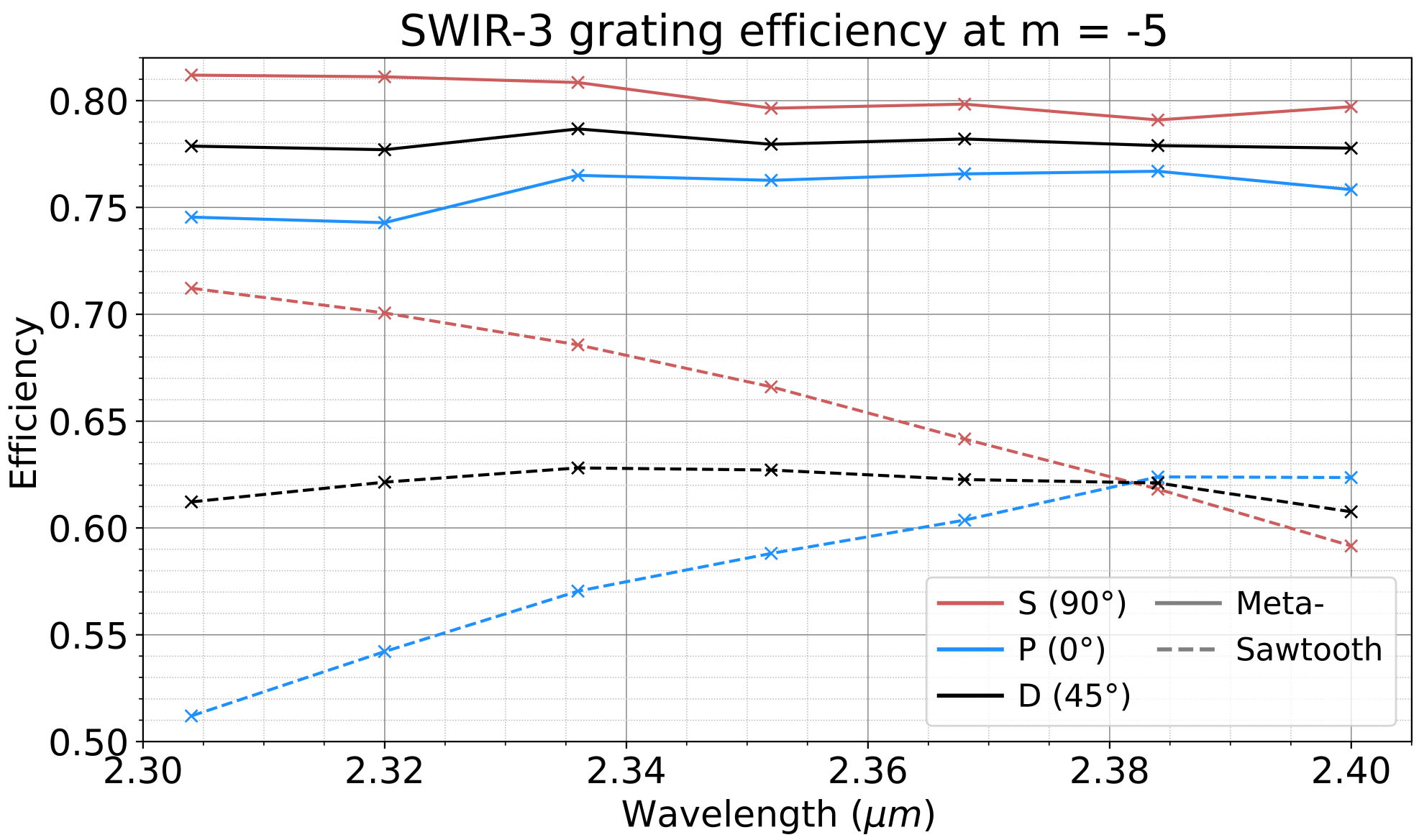

Fig. 5. Grating efficiency as a function of wavelength range for the metagrating (in solid lines) and sawtooth (in dashed lines). for S-polarized light (red), P-polarized light (blue) and Diagonal polarized light (black). The wavelength at which the efficiency is simulated are shown in crosses. The efficiencies are at grating order (m=) $-5$ corresponding to a diffraction angle of $\sim 49.8° - 53.2°$ (depending on the wavelength). The incidence angle of the source used is 62.6°

The sawtooth grating shows a smooth, gradual spectral variation compared to the metagrating's staggered response. This smoothness is predicted from the scalar diffraction theory. According to that, an ideal blazed grating reaches peak diffraction efficiency when the operating wavelength matches the blazing wavelength. And this efficiency follows a sinc-squared envelope that rapidly declines away from the blaze [5]. The reason for the metagrating's staggered spectral features is not entirely clear. One possible explanation is that the discrete binary structures of the metagrating introduce sharp transitions in the electromagnetic response due to localized resonances and abrupt geometry changes between neighbouring unit cells.

| | **Metagrating** | **Sawtooth grating** |
|---|---|---|
| **Polarization** | **Average efficiency** [%] | **Average efficiency** [%] |
| S | 80.2 | 65.9 |
| P | 75.8 | 58.1 |
| D | 78.7 | 62.0 |

Table 2. Comparison of averaged grating efficiencies over the SWIR-3 band, for the immersed metagrating and immersed sawtooth grating designs under different polarizations.

The polarization sensitivity, defined by Equation 3, is shown in Figure 6 across the SWIR-3 band. The figure compares the sensitivity of the sawtooth grating with that of the metagrating.

The metagrating maintains a sensitivity within 5%, whereas the sawtooth grating exhibits values approaching 15%. Notably, the sensitivity curve for the sawtooth crosses zero, indicating a shift in polarization preference across the band. In contrast, the metagrating consistently shows a positive sensitivity, indicating a slightly better performance for S-polarization.

The behaviour can be tailored based on design goals by appropriately defining the figure of merit (FOM) during the optimization process. That could be used to compensate for the polarization sensitivity from other components in a grating spectrometer.

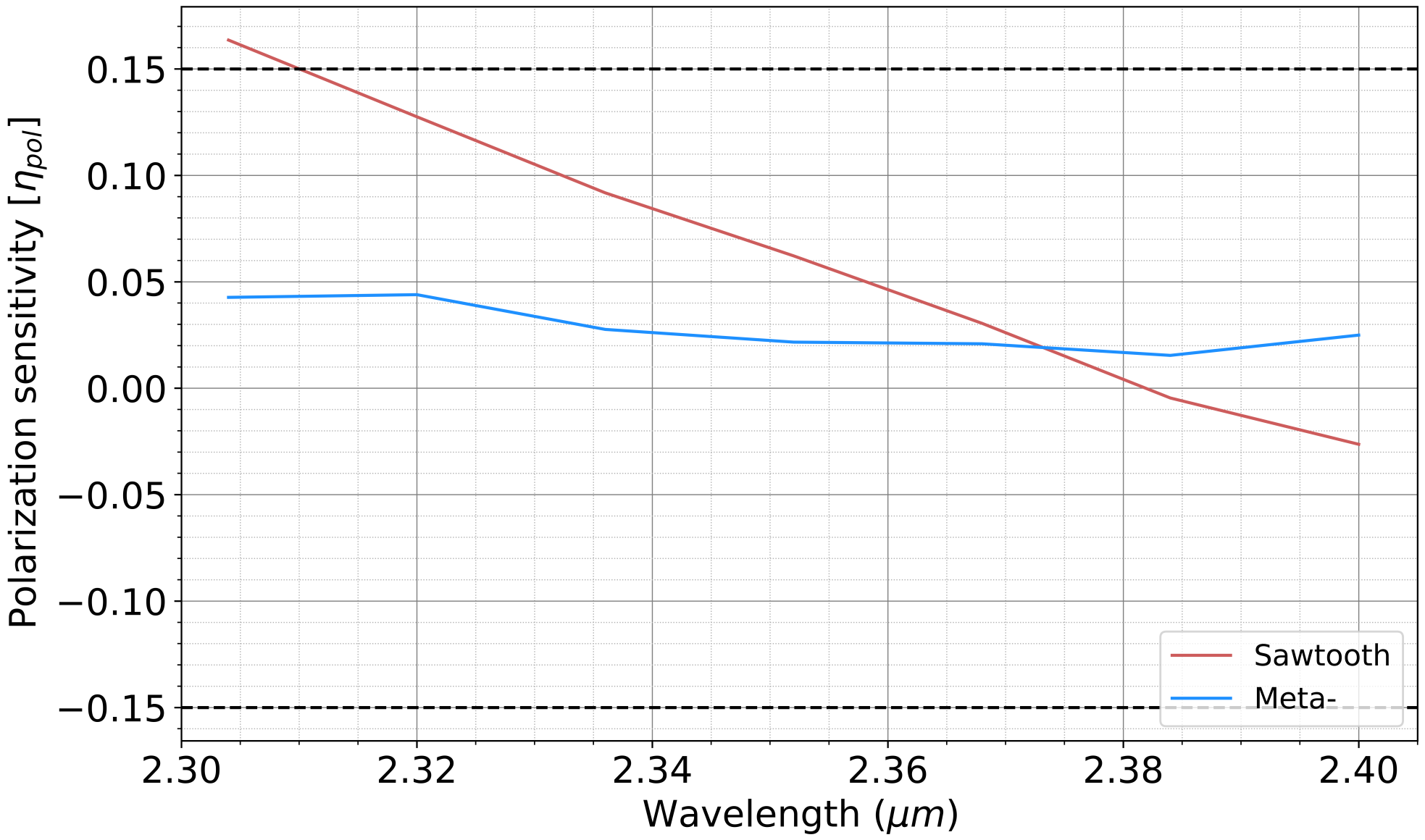

Fig. 6. Polarisation sensitivity ($\eta_{\text{pol}}$) as a function of the wavelength range for the immersed metagrating (in blue line) and immersed sawtooth grating (in red line). As a reference, the two black dashed lines mark $|\eta_{\text{pol}}| \leq 15$ %, which was the requirement for the Sentinel-5 mission.

Finally, Figure 7 shows the dispersion of the grating orders for the metagrating and compares it to that of the conventional sawtooth grating. The diffraction angles from the metagrating for all the orders match those of the sawtooth grating. Zooming in on the most excited −5th order in Figure 7 (a), it can be noticed that the diffraction angles for the sawtooth and metagrating are in good agreement.

The performance of the metagrating was evaluated in comparison to the conventional sawtooth design, particularly in terms of diffraction efficiency, polarization insensitivity, and chromatic stability. Moreover, since the metagrating is a flat periodic element, it doesn't introduce any aberrations when its far-field pattern is simulated, and the resulting point source function is diffraction-limited. While these results demonstrate the effectiveness of the metagrating, it is equally important to understand the underlying physical mechanisms that enable such performance. A perfectly implemented groove shape of a blazed grating in one period introduces an optical path difference of roughly one wavelength, i.e., the wavefront acquires a 0 to $2\pi$ phase response (which could be wrapped multiple times within one period). Hence, the phase response of the immersed metagrating is studied to further understand its performance.

The weighted wrapped phase profiles of the diffracted fields for the metagrating and the sawtooth grating are shown in Figure 8. Figures 8(a) and (b) display the phase of the $E_x$ and $E_z$ components, respectively, over one period of the metagrating for S- and P-polarized incident light, while (c) and (d) show the $E_x$ and $E_y$ components for the sawtooth grating. To prevent unphysical phase fluctuations in regions with very low field magnitude, the weak components were weighted by their amplitude. For example, under S-polarized input, the $E_x$ component carries little energy and is thus amplitude-weighted to suppress noise in the phase. Conversely, for the strong component—the $E_z$ component under S-polarization and the $E_x$ component under P-polarization—the field was left unweighted (i.e., multiplied by 1).

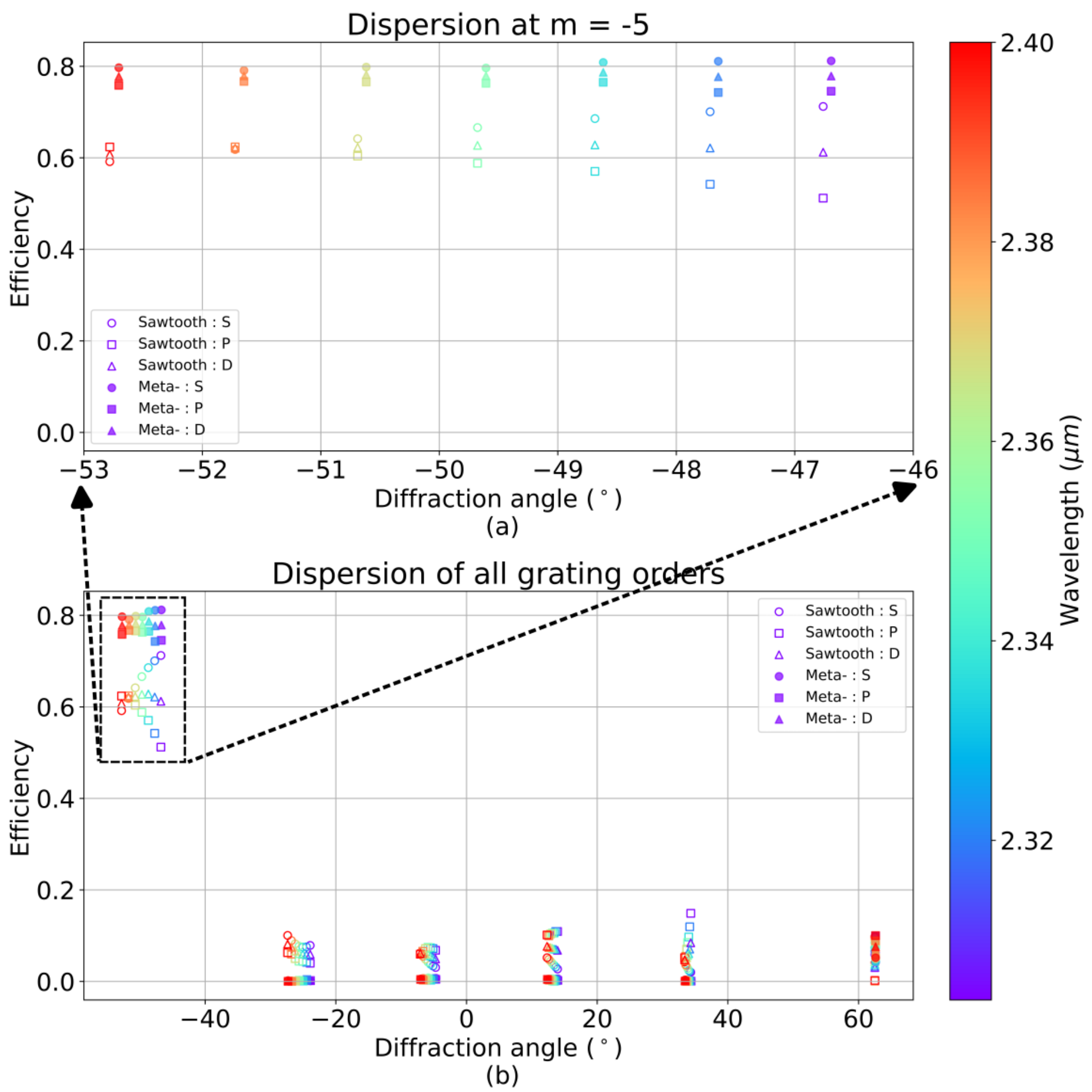

Fig. 7. Dispersion performance of the metagrating compared with the sawtooth grating. The efficiency as a function of the diffraction angle, with the color of the markers indicating the wavelength using the colorbar. The six clusters of points in figure (b) show the six excited orders 0, -1, -2, -3, -4, -5 from left to right, respectively. A zoomed-in region around the grating order of interest (-5) is shown in figure (a). Markers with face color are for metagrating, and those without face color are for sawtooth. Each different shape, circles, squares and triangles, corresponds to a different polarization, S, P, and D, respectively.

The phase profiles for the metagrating exhibit smooth variation, forming a clear spatial phase gradient from $-\pi$ to $\pi$ across the grating period. Such a phase ramp response over one period of the grating explains the ability to concentrate light in a particular diffraction order. The consistent shape of the curves across wavelengths (as indicated by the color map) confirms the robust performance of the structure, with minimal chromatic phase distortion. Furthermore, the similar phase ramps for both S- and P-polarizations indicate that the metagrating achieves its low polarization sensitivity by producing nearly identical effective phase gradients for both input states. In comparison, the sawtooth grating exhibits the smooth $-\pi$ to $\pi$ phase response over one grating period only at a few wavelengths, which explains the degradation of its efficiency.

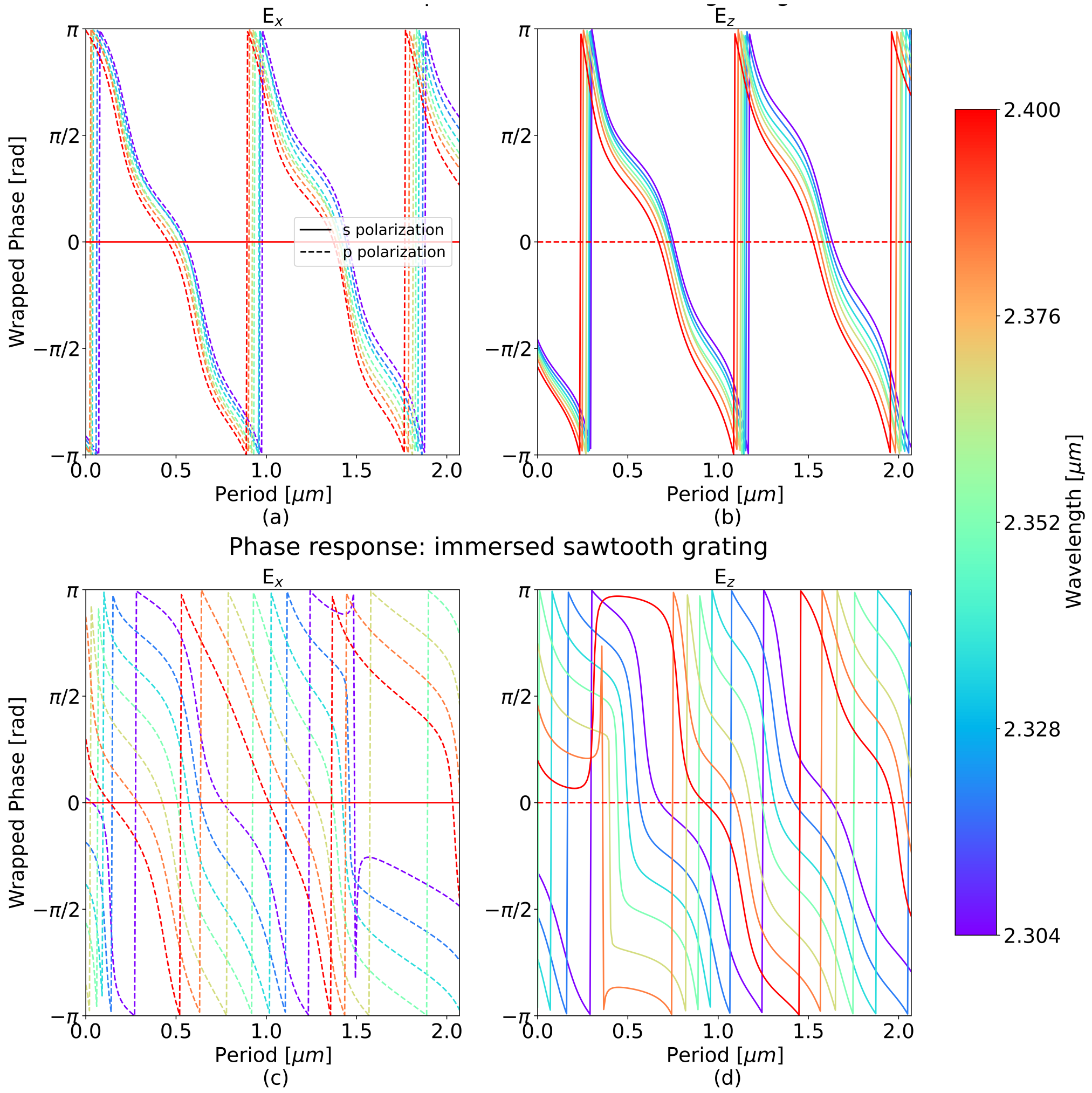

Fig. 8. Phase response of the metagrating and sawtooth grating across the SWIR-3 band. Figures (a) and (c) shows the wrapped phase of the $E_x$ component for meta and sawtooth grating respectively. Figures (b) and (d) shows the wrapped phase of the $E_z$ component of the diffracted light for meta and sawtooth grating respectively. The profiles are plotted over one period of the grating ($P = 2.07\,\mu$m). Solid lines correspond to S-polarized input and dashed lines to P-polarized input. Each color indicates a different wavelength shown in the colorbar. For components with negligible field strength ($E_x$ under S-polarization and $E_z$ under P-polarization), the phase is amplitude-weighted to suppress noise in low-intensity regions.

The phase profiles for the immersed metagrating over the grating period can also be interpreted as the cumulative phase response of individual metasurfaces over one period. Hence, the grating response can be optimized by performing individual metasurface simulations for the appropriate phase and maximum transmission, thereby realizing the same phase response.

### *3.3. Manufacturing considerations*

As mentioned earlier, the feature size (> 150 nm) and the small aspect ratio (~ 1.3) make this design feasible using photolithography and have been demonstrated with even smaller feature sizes (< 150 nm) [38–41]. Photolithographic fabrication is inherently susceptible to systematic errors, including focus offsets, feature size variations, rounding artifacts, etch-depth inaccuracies, and mask misalignment [42]. Assessing the tolerance of the proposed design to such imperfections is therefore essential. Systematic errors are typically on the order of ~ ±1–5 nm. In the present study, the design performance was evaluated under a systematic deviation $\Delta\mathbf{x}$ from the optimal parameters $\mathbf{x}$ by computing the grating efficiency for $\mathbf{x} \pm \Delta\mathbf{x}$. The analysis was carried out until a performance degradation of approximately 10% was observed, which is taken as the effective tolerance limit.

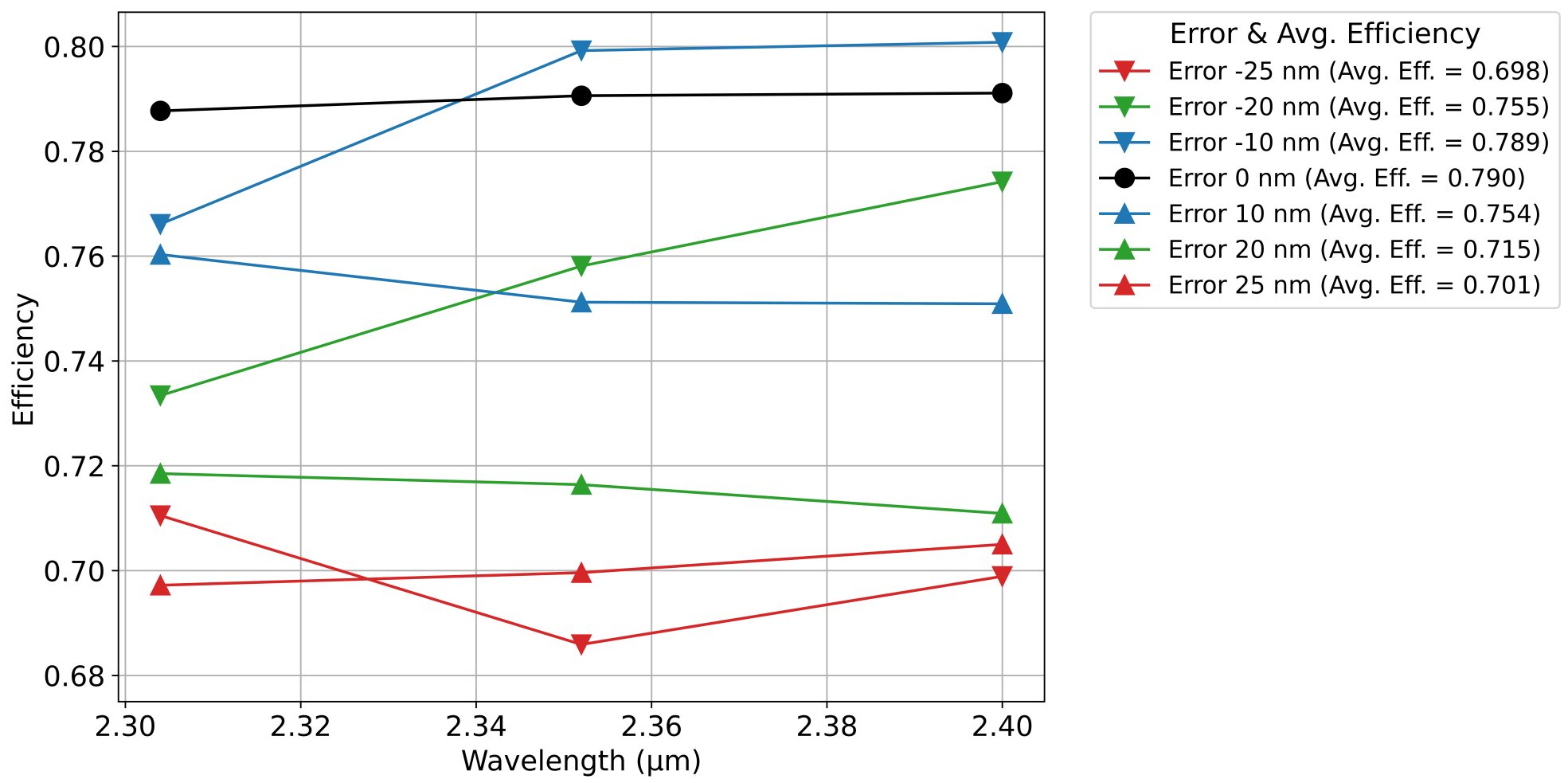

Fig. 9. Tolerance of the metagrating to the systematic error in the optimized parameters, during manufacturing. The tolerance is shown via the grating efficiency of the metagrating at average polarization (D) over the SWIR-3 band. The scatter point color represents the absolute error (in nm), while the marker shape indicates the sign of the error (upward triangle = positive error, downward triangle = negative error, circle = zero error). The points are shown at the minimum, maximum and the average of the SWIR-3 band since those points were used during optimization. The legend box also shows the average efficiency across the band for each of the systematic error cases.

Figure 9 shows the efficiency at D-polarization over the spectral band for systematic errors of ± (25, 20, 10) nm, compared with the optimal case. As expected, the ±25 nm deviation results in the largest reduction, with the efficiency degrading by approximately 10%, which defines the effective tolerance limit of the design. Although the −10 nm case may appear to exceed the nominal (0 nm) case at certain wavelength points, the average efficiency over the spectral range (evaluated at three wavelength points) is nearly the same in both cases: ~ 79.0% for the nominal case and ~ 78.9% for the −10 nm case. The results also indicate that negative and positive deviations of the same magnitude affect efficiency differently, reflecting an asymmetry in the device's tolerance response. A general trend is also observed in which negative dimensional errors yield slightly higher average efficiency than positive errors of the same magnitude. The origin of this asymmetry is not fully clear and may be related to subtle variations in the effective phase and transmission response of the metasurface elements under over- and under-systematic error conditions. Even better tolerance can be expected with sub-nm systematic errors achievable with e-beam lithography when more precision and accurate performance is required.

## 4. Summary and outlook

An immersed metagrating was designed with fixed geometric constraints ($P = 2.07\,\mu$m, $\theta_i = 62.6°$, diffraction order $m = -5$, $\theta_d = -49.8°$), derived from the Sentinel-5 SWIR-3 grating requirements ($\lambda_{\text{mean}} = 2.345\,\mu$m, $\sim 4\%$ bandwidth). To remain within fabrication limits while satisfying the subwavelength condition, the grating period was discretized into five sub-periods (or lattice constant) ($\Lambda = 0.414\,\mu$m). The Rytov-based EMT approximations can effectively estimate refractive indices but cannot guarantee optimal efficiency in this regime, particularly given the finite extinction coefficients of Si and Al. Additionally, to use the Rytov-based EMT to approximate a profile, an ideal profile is required that performs well across the spectral range and accounts for polarization effects. Therefore, the design problem was formulated as a black-box optimization with eleven degrees of freedom (5 widths, 5 lengths, and 1 height) and solved using modified Covariance Matrix Adaptation Evolution Strategy (CMA-ES). The figure of merit (FOM) was defined as the average $-5^{\text{th}}$ order grating efficiency at diagonal polarization across three representative wavelengths (minimum, mean and maximum) of the interested spectral band.

The CMA-ES optimization converged within $\sim 500$ evaluations to a stable set of parameters (Table 1), with pillar dimensions above 150 nm and aspect ratios $\sim 1.3$, compatible with both e-beam and photolithography. The optimized design achieved grating efficiencies above 74% across the SWIR-3 band, with average values of 80.2% (S), 75.8% (P), and 78.7% (D), representing an improvement of $\sim 15\%$ compared to a conventional immersed blazed grating designed for the Sentinel-5 mission (Table 2). Polarization sensitivity was reduced to $< 5\%$ across the band (Figure 6). The metagrating also demonstrated reduced chromatic variation, with nearly flat efficiency curves across wavelength, and diffraction angles deviating from the conventional sawtooth case by less than 0.08° (Figure 7). Phase analysis confirmed the presence of smooth $-\pi$ to $\pi$ ramps across one grating period, nearly identical for S and P polarizations (Figure 8), explaining the improved efficiency and reduced polarization dependence.

Finally, tolerance studies showed that systematic dimensional errors of $\pm 25$ nm reduce average efficiency by $\sim 10\%$, while $\pm 10$ nm deviations resulted in changes $< 1\%$ (Figure 9). The asymmetry between negative and positive deviations suggests a nontrivial dependence of efficiency on feature size perturbations, but overall robustness to fabrication variability was confirmed. These results demonstrate that the immersed metagrating design not only outperforms the conventional sawtooth blazed grating in terms of efficiency and polarization insensitivity, but also remains feasible with current lithographic manufacturing processes. Furthermore, with advanced lithographic techniques, the same process could be applied to design gratings for shorter wavelengths (optical and UV). However, such implementations would have to rely solely on e-beam lithography, since the minimum feature size of the metagrating would fall below 100 nm. This arises because the minimum feature size of metasurfaces scales with the grating period, which in turn scales with the wavelength. Consequently, metagrating designs are more readily realizable for mid-IR or longer wavelengths, where they can also improve the performance of gratings currently achievable.

For this study, a more conservative estimation was made to keep the manufacturing of the designed metagrating more achievable. The restriction on the number of divisions per period (here, 5) can be relaxed and could be made a free parameter for optimization. This would be especially beneficial due to advanced e-beam lithography techniques, where sub-10 nm feature sizes are possible [43]. Additionally, including more geometries in the optimization process—such as pillars with differently shaped bases, hollowed-out structures, and/or irregularly spaced meta-surfaces—could further help realize more control over the phase profile [44], which in turn would help to achieve the perfect blazed profile across a broader band over both polarizations.

It is also important to consider the limitations of the simulations. Some manufacturing defects are not included, such as imperfections in the aluminum coating/filling of the etched metasurfaces. Furthermore, wavefront errors and ghosts were not simulated, and therefore, in practice, the efficiency of the grating may slightly deviate from the simulations [27]. The impact of such imperfections can be best assessed by manufacturing the metagrating. Hence, the natural next step would be to test the designed metagrating by realizing it using either DUV lithography (for prototyping) or e-beam lithography (for more accurate realization).

**Data availability.** Data underlying the results and the code for the simulations presented in this paper are available at [45].

**Disclosures.** The authors declare no conflicts of interest.

## References

1. E. M. George, D. Gräff, H. Feuchtgruber, *et al.*, "Making spiffi spiffier: upgrade of the spiffi instrument for use in eris and performance analysis from re-commissioning," in *Ground-based and Airborne Instrumentation for Astronomy VI,* vol. 9908 (SPIE, 2016), pp. 123–142.
2. T. Agócs, E. Elswijk, D. Zaalberg, *et al.*, "Ge immersed grating manufacturing and optical verification for the metis high-resolution spectrograph," in *Advances in Optical and Mechanical Technologies for Telescopes and Instrumentation IV,* vol. 11451 (SPIE, 2020), pp. 306–317.
3. D. T. Jaffe, S. Barnes, C. Brooks, *et al.*, "Gmtnirs: progress toward the giant magellan telescope near-infrared spectrograph," in *Ground-based and Airborne Instrumentation for Astronomy VI,* vol. 9908 (SPIE, 2016), pp. 648–656.
4. R. Kohlhaas, P. Tol, R. Schuurhof, *et al.*, "Manufacturing and optical performance of silicon immersed gratings for sentinel-5," in *International Conference on Space Optics—ICSO 2018,* vol. 11180 (SPIE, 2019), pp. 585–605.
5. Y. Soskind, *Field guide to diffractive optics* (SPIE, 2011).
6. H. G. Philipsen and J. J. Kelly, "Anisotropy in the anodic oxidation of silicon in koh solution," The J. Phys. Chem. B **109**, 17245–17253 (2005).
7. N. Yu and F. Capasso, "Flat optics with designer metasurfaces," Nat. materials **13**, 139–150 (2014).
8. P. Berini, "Optical beam steering using tunable metasurfaces," Acs Photonics **9**, 2204–2218 (2022).
9. F. Ding, Z. Wang, S. He, *et al.*, "Broadband high-efficiency half-wave plate: a supercell-based plasmonic metasurface approach," ACS nano **9**, 4111–4119 (2015).
10. M. Erdmann, E.-B. Kley, and U. Zeitner, "Development of a large blazed transmission grating by effective binary index modulation for the gaia radial velocity spectrometer," in *International Conference on Space Optics—ICSO 2010,* vol. 10565 (SPIE, 2017), pp. 373–378.
11. Z. Peng, D. A. Fattal, A. Faraon, *et al.*, "Reflective silicon binary diffraction grating for visible wavelengths," Opt. letters **36**, 1515–1517 (2011).
12. A. Kitt, J. Rolland, and A. Vamivakas, "Visible metasurfaces and ruled diffraction gratings: a comparison," Opt. Mater. Express **5**, 2895–2901 (2015).
13. P. Lalanne, S. Astilean, P. Chavel, *et al.*, "Blazed binary subwavelength gratings with efficiencies larger than those of conventional échelette gratings," Opt. letters **23**, 1081–1083 (1998).
14. M.-S. L. Lee, P. Legagneux, P. Lalanne, *et al.*, "Blazed binary diffractive gratings with antireflection coating for improved operation at 10.6 $\mu$m," Opt. Eng. **43**, 2583–2588 (2004).
15. M.-S. L. Lee, J. Cholet, A. Delboulbé, *et al.*, "Wide band uv/vis/nir blazed-binary reflective gratings for spectro-imagers: two lithographic technologies investigation," J. Eur. Opt. Soc. Publ. **19**, 7 (2023).
16. K. Zhang, Y. Wang, S. N. Burokur, and Q. Wu, "Generating dual-polarized vortex beam by detour phase: From phase gradient metasurfaces to metagratings," IEEE Trans. on Microw. Theory Tech. **70**, 200–209 (2021).
17. N. A. Rubin, A. Zaidi, M. Juhl, *et al.*, "Polarization state generation and measurement with a single metasurface," Opt. express **26**, 21455–21478 (2018).
18. U. D. Zeitner, H. Dekker, F. Burmeister, *et al.*, "High efficiency transmission grating for the eso cubes uv spectrograph," Exp. astronomy **55**, 281–300 (2023).
19. U. D. Zeitner, M. Oliva, F. Fuchs, *et al.*, "High performance diffraction gratings made by e-beam lithography," Appl. Phys. A **109**, 789–796 (2012).
20. P. Cheben, R. Halir, J. H. Schmid, *et al.*, "Subwavelength integrated photonics," Nature **560**, 565–572 (2018).
21. S. Rytov, "Electromagnetic properties of a finely stratified medium," Sov. Phys. JEPT **2**, 466–475 (1956).
22. H. Hemmati and R. Magnusson, "Applicability of rytov's full effective-medium formalism to the physical description and design of resonant metasurfaces," ACS Photonics **7**, 3177–3187 (2020).

23. I. K. Baldry, J. Bland-Hawthorn, and J. Robertson, “Volume phase holographic gratings: polarization properties and diffraction efficiency,” Publ. Astron. Soc. Pac. **116**, 403 (2004).
24. S. Ans, F. Zamkotsian, and G. Demésy, “Topology optimization of blazed gratings under conical incidence,” J. Opt. Soc. Am. A **41**, 1531–1543 (2024).
25. J. de Nobel, D. Vermetten, H. Wang, *et al.*, “Tuning as a means of assessing the benefits of new ideas in interplay with existing algorithmic modules,” in *Proceedings of the Genetic and Evolutionary Computation Conference Companion,* (2021), pp. 1375–1384.
26. J. P. Marsh, O. A. Ershov, and D. T. Jaffe, “Silicon grisms and immersion gratings produced by anisotropic etching: testing and analysis,” in *IR Space Telescopes and Instruments,* vol. 4850 J. C. Mather, ed., International Society for Optics and Photonics (SPIE, 2003), pp. 797 – 804.
27. U. Zeitner, D. Michaelis, E.-B. Kley, and M. Erdmann, “High performance gratings for space applications,” in *Micro-Optics 2010,* vol. 7716 (SPIE, 2010), pp. 438–445.
28. C. Audet and W. Hare, *Derivative-Free and Blackbox Optimization*, SpringerLink Bücher (Springer, Cham, 2017).
29. N. Hansen and A. Ostermeier, “Completely derandomized self-adaptation in evolution strategies,” Evol. computation **9**, 159–195 (2001).
30. N. Hansen, A. Auger, R. Ros, *et al.*, “Comparing results of 31 algorithms from the black-box optimization benchmarking BBOB-2009,” in *Proceedings of the 12th annual conference companion on Genetic and evolutionary computation,* (2010), GECCO ’10, pp. 1689–1696.
31. J. de Nobel, D. Vermetten, H. Wang, *et al.*, “Tuning as a means of assessing the benefits of new ideas in interplay with existing algorithmic modules,” in *Genetic and Evolutionary Computation Conference, GECCO ’21, Companion Volume, Lille, France, July 10-14, 2021,* K. Krawiec, ed. (ACM, 2021), pp. 1375–1384.
32. G. A. Jastrebski and D. V. Arnold, “Improving evolution strategies through active covariance matrix adaptation,” in *2006 IEEE international conference on evolutionary computation,* (IEEE, 2006), pp. 2814–2821.
33. A. Auger, D. Brockhoff, and N. Hansen, “Mirrored sampling in evolution strategies with weighted recombination,” in *13th Annual Genetic and Evolutionary Computation Conference, GECCO 2011, Proceedings, Dublin, Ireland, July 12-16, 2011,* N. Krasnogor and P. L. Lanzi, eds. (ACM, 2011), pp. 861–868.
34. H. Smaoui, L. Zouhri, S. Kaidi, and E. Carlier, “Combination of fem and cma-es algorithm for transmissivity identification in aquifer systems,” Hydrol. Process. **32**, 264–277 (2018).
35. B. Ghahremani, M. Bitaraf, and H. Rahami, “A fast-convergent approach for damage assessment using cma-es optimization algorithm and modal parameters,” J. Civ. Struct. Health Monit. **10**, 497–511 (2020).
36. H. Hamdani, B. Radi, and A. El Hami, “Optimization of solder joints in embedded mechatronic systems via kriging-assisted cma-es algorithm,” Int. J. for Simul. Multidiscip. Des. Optim. **10**, A3 (2019).
37. U. D. Zeitner, M. Banasch, and M. Trost, “Potential of e-beam lithography for micro-and nano-optics fabrication on large areas,” J. Micro/Nanopatterning, Materials, Metrol. **22**, 041405–041405 (2023).
38. T. Hu, Q. Zhong, N. Li, *et al.*, “Cmos-compatible a-si metalenses on a 12-inch glass wafer for fingerprint imaging,” Nanophotonics **9**, 823–830 (2020).
39. M. Keil, A. E. Wetzel, K. Wu, *et al.*, “Large plasmonic color metasurfaces fabricated by super resolution deep uv lithography,” Nanoscale advances **3**, 2236–2244 (2021).
40. V. Vakarin, D. Melati, T. T. D. Dinh, *et al.*, “Metamaterial-engineered silicon beam splitter fabricated with deep uv immersion lithography,” Nanomaterials **11**, 2949 (2021).
41. E. Khaidarov, D. Eschimese, K. H. Lai, *et al.*, “Large-scale vivid metasurface color printing using advanced 12-in. immersion photolithography,” Sci. Reports **12**, 14044 (2022).
42. J. Wang, F. Zhang, Q. Song, *et al.*, “Fabrication error analysis for diffractive optical elements used in a lithography illumination system,” Opt. Eng. **54**, 045102–045102 (2015).
43. J. Gour, S. Beer, P. Paul, *et al.*, “Wafer-scale nanofabrication of sub-5 nm gaps in plasmonic metasurfaces,” Nanophotonics **13**, 4191–4202 (2024).
44. S. So, J. Mun, J. Park, and J. Rho, “Revisiting the design strategies for metasurfaces: fundamental physics, optimization, and beyond,” Adv. Mater. **35**, 2206399 (2023).
45. D. Patel, J. de Nobel, and R. Kohlhaas, “metagratingModCMAES,” https://github.com/dhwanilpate1/metagratingModCMAES (2025).